\documentclass[journal=jctcce,manuscript=article]{achemso}
\usepackage[version=3]{mhchem} 
\usepackage[T1]{fontenc}       
\usepackage{braket}
\usepackage{amsmath}
\usepackage{amssymb}
\usepackage{mathtools}
\usepackage{color}
\usepackage[cdot,amssymb]{SIunits}
\usepackage[inline]{enumitem}
\usepackage{graphicx}
\usepackage[caption=false]{subfig}
\usepackage{hyperref}

\author{Jiajun Ren}
 \affiliation{MOE Key Laboratory of Organic OptoElectronics and Molecular
 Engineering, Department of Chemistry, Tsinghua University, Beijing 100084,
 People's Republic of China }

\author{Zhigang Shuai}%
\email{zgshuai@tsinghua.edu.cn}
 \affiliation{MOE Key Laboratory of Organic OptoElectronics and Molecular
 Engineering, Department of Chemistry, Tsinghua University, Beijing 100084,
 People's Republic of China }

\author{Garnet Kin-Lic Chan}
\email{gkc1000@gmail.com}
 \affiliation{Division of Chemistry and Chemical Engineering, California
 Institute of Technology, Pasadena, California 91125, United States}

\title{Time Dependent Density Matrix Renormalization Group Algorithms for Nearly
Exact Absorption and Fluorescence Spectra of Molecular Aggregates at Both Zero
and Finite Temperature} 

\begin{document}

\begin{tocentry}
      \centering
      \includegraphics[scale=0.33]{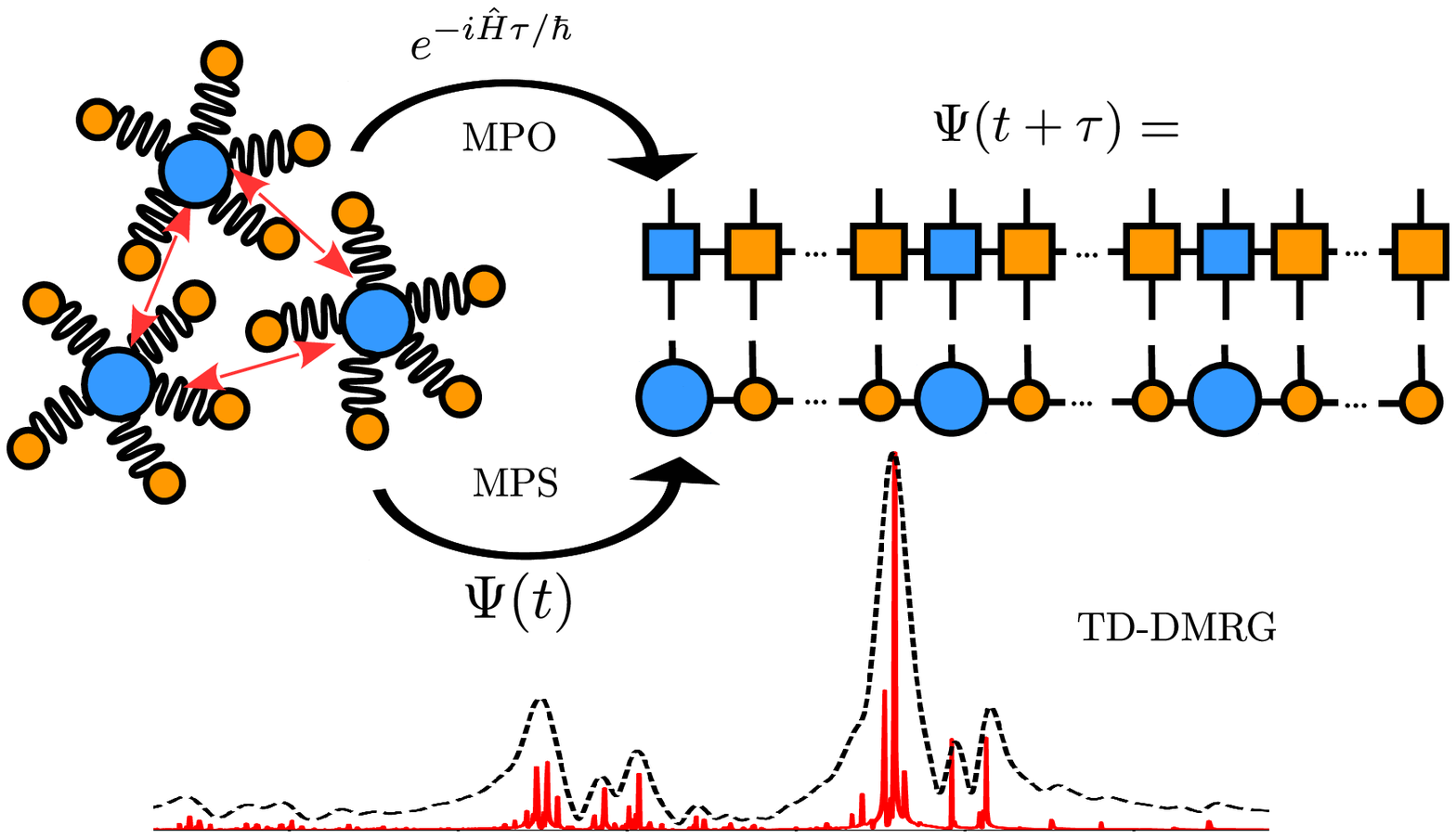}  

\end{tocentry}

\begin{abstract}
    We implement and apply time-dependent density matrix renormalization group
    (TD-DMRG) algorithms at zero and finite temperature to compute the linear
    absorption and fluorescence spectra of molecular aggregates.  Our
    implementation is within a matrix product state/operator framework with an
    explicit treatment of the excitonic and vibrational degrees of freedom, and
    uses the locality of the Hamiltonian in the zero-exciton space to improve
    the efficiency and accuracy of the calculations.  We demonstrate the power
    of the method by calculations on several molecular aggregate models,
    comparing our results against those from multi-layer multiconfiguration
    time-dependent Hartree and n-particle approximations.  We find that TD-DMRG
    provides an accurate and efficient route to calculate the spectrum of
    molecular aggregates.
\end{abstract}

\section{Introduction}
    The effects of aggregation on the optical and transport properties of
    molecular aggregates and polymers, such as aggregation induced emission
    (AIE) in the siloles \cite{luo2001aggregation} and ultrafast energy transfer
    in the photosynthetic light harvesting complexes (LHC),
    \cite{engel2007evidence} have attracted tremendous attention in the past
    decades.  Understanding the change in the linear spectrum moving from a
    single molecule to an aggregate is usually the first step to understand the
    effects of aggregation.  Theoretically, the quantitative calculation of the
    spectrum of molecular aggregates is challenging, as it is a problem of
    coupled many-particle quantum dynamics, including both excitonic coupling
    and exciton-vibrational (phonon) coupling. \cite{schroter2015exciton} Though
    perturbative methods are very successful in treating the two limiting
    coupling regimes \cite{jang2004multichromophoric, schroder2006calculation}
    --- the weak excitonic coupling regime and the weak exciton-vibrational
    coupling regime --- many interesting systems, such as the LHC, lie in the
    intermediate coupling regime, and remain challenging to model accurately,
    motivating the development of a wide variety of approximate modeling
    methods. \cite{tanimura1989time, chen2009optical, diosi1997non, roden2011non,
    ke2017hierarchy, ke2017extension, makri1995tensor, philpott1967vibronic,
    hoffmann2002optical, zhao2005vibronic, gao2011vibronic, meyer2009multidimensional}
    
    Current numerically exact approaches for the intermediate coupling regime
    work within one of two representations. The first representation is most
    closely associated with the theory of open quantum system dynamics and
    dissipative dynamics. Here the whole problem is divided into a system
    (usually containing the electronic degrees of freedom) and an environment
    (usually containing the vibrational degrees of freedom). Then, an effective
    equation of motion for the system part only is derived by eliminating the
    environmental degrees of freedom, assuming a linear system-environment
    coupling, and with the environment typically approximated as an infinite
    harmonic bath described by a continuous spectral density. Some
    representative numerically exact methods using this representation include
    the hierarchical equation of motion method (HEOM), \cite{tanimura1989time,
    chen2009optical} stochastic Schr\"{o}dinger equation
    (SSE),\cite{diosi1997non, roden2011non} hybrids of HEOM and
    SSE,\cite{ke2017hierarchy, ke2017extension} and the quasi-adiabatic
    propagator path-integral (QUAPI) method~\cite{makri1995tensor}. In the
    second representation, the whole problem is treated explicitly as a closed
    finite-dimensional system, with discrete vibrational degrees of freedom.
    Some numerically exact methods commonly employed for such problems include
    exact diagonalization (ED) and the multi-configuration time dependent
    Hartree (MCTDH) method.  \cite{meyer2009multidimensional} 

    The density matrix renormalization group (DMRG) methods we use in this work
    are formulated within the second representation, and thus the most relevant
    methods to compare against are ED and MCTDH. Full ED, whether applied to
    the zero temperature linear spectrum \cite{gagliano1987dynamical} or within
    the more recent finite temperature Lanczos formulation
    \cite{kokalj2009finite, prelovvsek2013ground} is limited to very small
    systems, as the dimension of the Hilbert space increases exponentially with
    system size.  To overcome the exponential wall in full ED, systematically
    improvable approximations can be introduced to reduce the cost, akin to the
    configuration interaction hierarchy in quantum chemistry
    \cite{szabo1996modern, zgid2012truncated}. One popular example is the
    so-called ``n-particle approximation'',\cite{philpott1967vibronic,
    hoffmann2002optical, zhao2005vibronic, gao2011vibronic} in which only the
    electronically excited molecule and at most n-1 nearest ground state
    molecules are allowed to be simultaneously vibrationally excited.  The
    motivation for this restricted Hilbert space is that a local exciton should
    only disturb phonon modes nearby, creating an ``exciton polaron''.
    \cite{spano2009spectral} Note that the n-particle approximation is also used
    outside of the ED context, for example, it has recently been used in HEOM to
    improve efficiency.\cite{song2015time} Though the n-particle approximation
    method has been applied to explain many features of the spectra of
    conjugated polymers and molecular aggregates,\cite{spano2006excitons,
    spano2009spectral} the lowest level 2-particle approximation is still too
    expensive to treat  intermediate sized aggregates with multiple phonon
    modes, such as multiple molecules with more than ten modes per molecule, the
    latter being essential to obtain fine structure in the
    spectrum.\cite{zhao2005vibronic} MCTDH is an alternative numerically exact
    method, based on a low rank tensor representation of the time-dependent
    wavefunction.  Its generalization --- multi-layer MCTDH (ML-MCTDH) --- has
    become a popular approach to numerically converge towards the exact quantum
    dynamics of systems  with dozens or even hundreds of quantum degrees of
	freedom.  \cite{wang2003multilayer,worth2008using} 
    Two algorithms have previously been described to include  finite temperature effects within MCTDH.  The first is based on
    statistical sampling of wavefunctions from an initial thermal ensemble.\cite{matzkies1999accurate, manthe2001partition} This approach is
    numerically efficient in cases when convergence can be reached without too many samples, for example at low temperatures. However, the
    scale of temperature is defined relative to the frequencies in the problem, thus, in systems with low frequency modes, where high quanta
    states would be populated even at room temperature, such an algorithm is not very efficient.  The second directly propagates the density
    matrix of the mixed state by Liouville's equation, and is called $\rho$-MCTDH. It was first explored by Raab \textit{et al} in
    1999,\cite{raab1999multiconfiguration, raab2000numerical} and is based on the finite temperature time dependent variational principle
    (TDVP).  Unlike  the first approach,  $\rho$-MCTDH does not introduce statistical error.  However, to our knowledge, the algorithm has
    not been widely used. One drawback is that propagating the density matrix is more expensive than propagating the wavefunction. Another
    drawback is that, unlike at zero temperature, the finite temperature TDVP that was used violates the conservation of the total energy and density
    matrix trace.\cite{mclachlan1964variational}

    In the current work, we consider an alternative numerical approach to obtain
    near-exact zero-temperature and finite-temperature spectra, via the
    time-dependent density matrix renormalization group (TD-DMRG). Much like
    ML-MCTDH, the DMRG is based on a low-rank tensor representation of the
    wavefunction.  It was originally proposed by White to treat one-dimensional
    strongly correlated systems, where it has become the method of choice to
    compute low-lying eigenstates.  \cite{White1992DMRG, white1993density} The
    DMRG was subsequently extended to frequency-dependent dynamic properties,
    via the Lanczos DMRG,\cite{hallberg1995density} correction-vector DMRG
    \cite{shuai1998linear} and the dynamical DMRG
    methods.\cite{jeckelmann2002dynamical} These closely related algorithms all
    provide dynamic properties at zero temperature in the frequency domain. DMRG
    algorithms have also been formulated at finite temperature.  The earliest
    attempts used renormalization of the transfer matrix (TMRG),
    \cite{nishino1995density,bursill1996density,naef1999autocorrelations} but
    more systematic finite temperature formulations emerged out of
    time-dependent density matrix renormalization group (TD-DMRG)
    algorithms\cite{cazalilla2002time,luo2003comment, vidal2004efficient,
    white2004real, daley2004time, feiguin2005time,  tdvpchan, haegeman2011time,
    haegeman2016unifying, dorando2009analytic}, which also provide a route to
    real-time zero-temperature dynamical properties. In particular, the thermal state can
    be obtained by imaginary time-propagation (either within a purified state
    formalism or via propagation of operators ) and subsequent propagation along
    the real-time axis then allows for the computation of finite-temperature
    dynamical quantities. \cite{verstraete2004matrix,zwolak2004mixed,
    feiguin2005finite, barthel2013precise}

    DMRG methods have been applied to a wide variety of problems in chemical
    physics.  Semi-empirical DMRG and ab initio DMRG methods have been developed
    in the last two decades for molecular quantum chemistry
    problems~\cite{shuai1997quantum, Chan2011QCDMRGrev} including for the
    calculation of dynamical
    properties~\cite{dorando2009analytic,ronca2017time}. DMRG methods have also
    been applied to electron-phonon problems, including the spin-Peierls model,
    \cite{caron1996density} and to the single mode Holstein model.
    \cite{jeckelmann1998density, tozer2014localization, barford2016polarons}
    Finally, TD-DMRG methods have been recently applied to open quantum system
    dynamics, both in the context of impurity problems,
    \cite{garcia2004dynamical, ganahl2015efficient, wolf2014solving,
    hallberg2015state} as well as traditional system-bath
    models\cite{prior2010efficient,chin2013role,yao2013dynamics, YaoYaoCS,
    YaoYaoCS2}.

    In this work, we develop TD-DMRG algorithms for zero- and finite-temperature
    dynamic properties of electron-phonon coupled systems, as applied to the
    linear absorption and fluorescence spectra of electron-phonon coupled
    molecular aggregates.  As a non-perturbative method, we will show that
    these algorithms allow us to compute accurate spectra across any range of
    coupling strengths. Furthermore, we will demonstrate that the
    high-efficiency of the TD-DMRG algorithm  allows us to calculate the
    spectrum for a system as large as an 18 monomer distyryl benzene (DSB)
    aggregate, including the fine structure from the phonon modes.
    
    The remaining sections of this paper are arranged as follows. In Section
    \ref{sec:theory} we first define the Hamiltonian of the exciton model used
    in this work. Next, we describe  the zero and finite temperature TD-DMRG
    methods used to calculate the absorption and fluorescence spectra and
    provide some computational details. In Section \ref{sec:results}, the
    absorption and fluorescence spectra of both model and real systems are
    calculated and compared to that from ML-MCTDH, the n-particle approximation,
    and experimental spectra. Finally, we present our conclusions in Section
    \ref{sec:conclusions}.

    \section{Theory}
    \label{sec:theory}
    \subsection{Model Hamiltonian}

    To map the spectral problem for a set of molecular aggregates to a
    practically solvable model, we first make several (reasonable)
    approximations:
    \begin{enumerate}[label=(\roman*)]
        \item The electronic excited states of a molecular aggregate are a linear
            combination of the local excited states of a single molecule.
            \label{itm:a1}
        \item The motion of the nuclei can be described as a collection of
            independent harmonic vibrations.  \label{itm:a2}
        \item Only intra-molecular vibrations are considered to be linearly coupled to the local electronic state, where the frequency
            $\omega$ of each vibration is the same for both the ground and excited states. \label{itm:a3}
    \end{enumerate}
    The combination of the second and third approximations is usually referred
    to as the ``displaced harmonic oscillator'' approximation. Within these
    approximations, we can map the Hamiltonian of a molecular aggregate to a
    multi-mode Holstein Hamiltonian, sometimes called the ``Frenkel'' or
    ``Frenkel-Holstein'' Hamiltonian. In addition, we can add additional
    processes to the Hamiltonian, relaxing the above three assumptions, such as
    inter-molecular charge transfer states\cite{merrifield1961ionized,
    hoffmann2003mixing, gao2011vibronic}, anharmonic vibrational effects
    \cite{chorovsajev2017temporal}, and inter-molecular electron-phonon
    interactions (Peierls terms). For simplicity, however, we will only consider
    the Holstein Hamiltonian in this work.
    
    Choosing  harmonic oscillator wavefunctions within the ground state potential
    energy surface as the vibrational basis, the Hamiltonian can be formulated in
     second quantization, 
    \begin{gather}
        \hat{H} = \sum_i \varepsilon_i a_i^\dagger a_i + 
        \sum_{ij} J_{ij} a_i^\dagger a_j \nonumber \\ 
        + \sum_{in} \omega_{in} b_{in}^\dagger b_{in}
        + \sum_{in} \omega_{in} g_{in} a_i^\dagger a_i (b_{in}^\dagger + b_{in}) 
        \label{eq:discreteH}
    \end{gather}
    $\varepsilon_{i}$ is the local excited state energy of molecule $i$;
    $J_{ij}$ is the excitonic coupling between molecule $i$ and $j$;
    $\omega_{in}$ and $g_{in}$ are respectively the harmonic frequency and
    electron-vibrational coupling parameter of normal mode $n$ of molecule $i$.
    Another alternative parameter frequently used to represent the strength of
    electron-vibrational coupling is Huang-Rhys factor $S$, where $S=g^2$.  All the
    parameters can be obtained from ab-initio quantum chemistry calculations of
    monomers and dimers, or fitted to experimental data. The vibrational modes
    in eq.  \eqref{eq:discreteH} are considered to be discrete, as is
    appropriate for single molecules or small aggregates.  In contrast, in the
    condensed phase, one often describes the behaviour of the
    macroscopic bath in terms of a continuous spectral density function,
    \begin{gather}
        J_i (\omega) = \pi \sum_{n} (g_{in} \omega_{in})^2
        \delta(\omega-\omega_{in})
    \end{gather}
    To use the methods in this work with such a formulation, one has to first
    discretize the spectral density. There are many such discretization
    algorithms that have been explored in studies of quantum impurity problems
    and to describe open system quantum dynamics. \cite{de2015discretize,
    chin2010exact, shenvi2008efficient, metzner1989correlated, wolf2014solving}
    We will only consider the discrete modes associated with molecular
    aggregates here, and thus we will not dive further into the details of
    discretizing a continuous bath.
    
\subsection{MPS, MPO, ZT-TD-DMRG and FT-TD-DMRG}
    The theory  of DMRG and its associated algorithms has been described in
    detail in many excellent reviews. \cite{Schollwock2011DMRGrev} Here, to be
    self-contained, we only briefly summarize the most essential parts. 
    \subsubsection{MPS and MPO}
\begin{figure*}[htbp]
    \centering
    \includegraphics[width = 0.7 \textwidth]{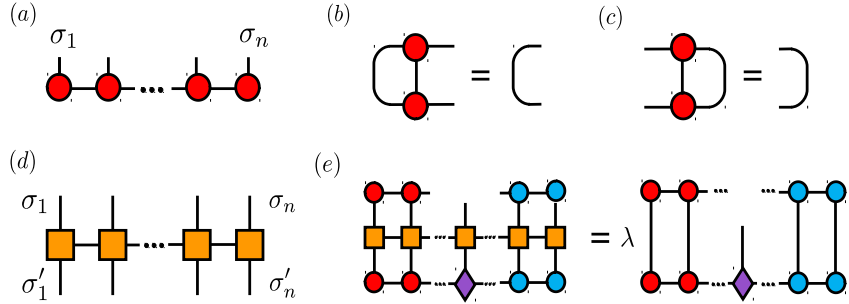}

    \caption{\label{fig:MPSblock}
        The graphical representation of (a) an MPS for $\Psi$ 
        (b) the left orthogonality condition (c) the right orthogonality condition
        (d) an MPO for $\hat{O}$
        (e) the ground state eigenvalue equation for $\hat{H}\psi = \lambda \psi$.
        Each linked bond represents a tensor contraction.}
    \end{figure*}

    DMRG is a wavefunction theory, where the many-body wavefunction amplitudes
    are obtained by multiplying out a chain of local matrices (thus giving rise
    to the moniker, matrix product state (MPS)),
    \begin{align}
        | \Psi \rangle & = \sum_{\{\sigma\}} C_{\sigma_1\sigma_2\cdots\sigma_n } | \sigma_1\sigma_2\cdots\sigma_n
        \rangle \nonumber \\
        & = \sum_{\{a\},\{\sigma\}}
         A^{\sigma_1}_{a_1} A^{\sigma_2}_{a_1a_2} \cdots
               A^{\sigma_n}_{a_{n-1}}  | \sigma_1\sigma_2\cdots\sigma_n \rangle
    \end{align}
    $a_n$ is the virtual bond index, whose ``bond dimension'' controls the
    accuracy of a DMRG calculation and the amount of entanglement captured by
    the MPS. $\sigma_n$ is called the physical bond index, and indexes the
    degrees of freedom of a single site. The MPS can be represented graphically
    as shown in Figure \ref{fig:MPSblock}(a).  There is a gauge freedom
    (redundancy in parametrization) in the MPS which can be fixed by requiring
    the matrices to the left or right of any given bond to satisfy respectively
    the left and right orthogonality conditions, $\sum_{\sigma}
    \mathbf{A}^{\sigma\dag} \mathbf{A}^\sigma  = \mathbf{I}$, $\sum_{\sigma}
    \mathbf{A}^{\sigma} \mathbf{A}^{\sigma \dag}  = \mathbf{I}$ respectively
    (see Figure \ref{fig:MPSblock}(b)(c)).
    
    The structure of the MPS wavefunction is related to that of the ML-MCTDH
    wavefunction. In particular, the ML-MCTDH wavefunction can be thought of as
    a tree factorization of the wavefunction amplitudes, also known as a tree
    tensor network state~\cite{shi2006classical}. Thus viewed from the ML-MCTDH
    perspective, the linear chain of matrices in the MPS formally corresponds to
    a maximally unbalanced tree, although such a tree structure is seldom used
    in ML-MCTDH. An advantage of the MPS structure over a more  general tree is
    that almost all numerical operations can be implemented as simple matrix
    operations rather than more general tensor contractions. In practical applications, this
    can lead to higher efficiency~\cite{nakatani2013efficient}.

    In the Holstein model we consider, there are both exciton and vibrational
    physical degrees of freedom. We associate each exciton and each vibration
    with its own matrix $\mathbf{A}^\sigma$ where $\sigma$ indexes the exciton
    or vibrational state. The exciton site has two states, $| 0 \rangle$
    representing the electronic ground state and $| 1 \rangle$  representing the
    electronic excited state. The vibrational site has $p$ states, $| 0
    \rangle, | 1 \rangle, \cdots | p-1 \rangle$ representing the number of
    phonons (in principle $p$ is infinite, but at low temperature, we can converge 
    static and dynamic properties with a finite $p$). In addition, the MPS is not
    invariant to the order in which the degrees of freedom are treated. We order
    the exciton and vibration degrees of freedom by molecule, that is  $e_1$,
    $\nu_{11}$, ..., $\nu_{1n}$, ..., $e_i$, $\nu_{i1}$, ..., $\nu_{in}$,
    ...($e_i$ is the exciton site of molecule $i$, $\nu_{in}$ is $n$th mode of
    molecule $i$).

    When constructing MPS algorithms, it is convenient to consider a related
    factorization of operators.  There, operator matrix elements are obtained by
    multiplying out a chain of local matrices, giving rise to a matrix product
    operator (MPO),
    \begin{gather}
        \hat{O} = \sum_{\{a\},\{\sigma\},\{\sigma'\}}
         W^{\sigma_1, \sigma'_1}_{a_1} W^{\sigma_2, \sigma'_2}_{a_1a_2} \cdots
                        W^{\sigma_n, \sigma'_n}_{a_{n-1}} \\ \nonumber |
                        \sigma_1\sigma_2\cdots\sigma_n \rangle \langle
                        \sigma'_1\sigma'_2 \cdots \sigma'_n |
    \end{gather}
    Similar to the gauge freedom in an MPS, there are many choices of local
    matrices in the MPO which multiply out to the same total operator $\hat{O}$.
    \cite{chan2016matrix} One optimal way to construct the MPO of the Holstein
    Hamiltonian in our work is given in Appendix \ref{appendix1}.

    To approximate a quantum ground state, we optimize the MPS matrices
    following the variational principle. Rather than optimize all matrices
    simultaneously, one typically optimizes the matrices one at a time, and this
    is the basis of the one-site DMRG algorithm. At each optimization step, the
    sites are partitioned into two blocks, A and B, and  an eigenvalue equation
    is solved in a space spanned by the direct product of the A and B block
    basis states $\ket{l}_A \otimes \ket{r}_B$, defined by the MPS matrices to
    the left and right of the partitioning bond respectively. If the gauge has
    been fixed such that sites in block A and block B satisfy left and right
    orthogonality conditions, then the eigenvalue equation takes the form
    \begin{gather}
        \sum_{l'r'} H_{lrl'r'} \psi_{l'r'} = E \psi_{lr} \label{eq:eigen}
    \end{gather}
    The quantities in the matrix eigenvalue equation and the block basis states
    are most easily visualized graphically (see Figure \ref{fig:MPSblock}(e)).
    The eigenvalue problem in eq. \eqref{eq:eigen} can be computed iteratively
    by the Davidson algorithm via series of tensor contractions with cost $O(2pM^3D
    + p^2M^2D^2)$ in each iteration, giving a total cost of $O(k_e k_\nu (2pM^3D
    + p^2M^2D^2))$ for one sweep of optimization steps over all sites. Here,
    $k_e$ is the number of molecules; $k_\nu$ is the number of normal modes of
    each molecule; $p$ is the number of local degrees of freedom; $M$ is the MPS virtual
    bond dimension; $D$ is the MPO bond dimension.  For the Holstein Hamiltonian,
    when considering only 1-dimensional nearest-neighbor excitonic couplings, $D
    \sim \textrm{const}$, otherwise $D \sim k_e$. 

    An important MPS algorithm is the compression of a large bond dimension MPS
    to a smaller bond dimension MPS. Compression is necessary because many
    algebraic operations involving MPS increase the bond dimension, such as
    acting an MPO on an MPS, or adding together two MPS (both operations are
    used in the time-dependent algorithms below). The simplest compression
    algorithm is the SVD compression algorithm.  Given the wavefunction
    $\psi_{lr}$ in \eqref{eq:eigen} on a given bond, we decompose it via SVD as
    \begin{align}
        \psi_{lr} \approx \sum_{d=1}^M s_d U_{ld} V^\dagger_{dr}
    \end{align}
    where $s_d$ are the singular values which can be truncated to some desired
    number $M$. Multiplying the truncated $\mathbf{U}$ and $\mathbf{V}$ matrices
    into the matrices to the left and right of the bond reduces the bond
    dimension joining the two matrices.  The compression can then be repeated
    for the next bond and iterated through the entire MPS. A closely related
    compression algorithm is the variational compression algorithm, where a
    compressed $\ket{\tilde{\psi}}$ with the desired reduced bond dimension is
    optimized to minimize the $L_2$ norm $\|\ket{\psi}-\ket{\tilde{\psi}}\|_2$,
    leading to a set of least squares equations to be solved at each site.  In
    this work, we employ the SVD algorithm for our MPS compressions.

    \subsubsection{ZT-TD-DMRG and FT-TD-DMRG}

    We now describe the zero-temperature and finite-temperature time-dependent
    DMRG algorithms.  For zero temperature TD-DMRG (ZT-TD-DMRG), we consider the
    initial state to be the ground-state MPS obtained via the above DMRG
    algorithm within a given exciton number sector (we track the exciton
    number as a good quantum number in the MPS sweep), i.e. the zero-(one-) exciton
    space for absorption(fluorescence).  We then need to propagate the state
    under the time independent Hamiltonian propagator $e^{-i\hat{H}t}$.  In the
    language of MPS and MPO, the task is to approximate the propagator as an MPO
    and the subsequent time-evolved state as an MPS in a computationally
    efficient manner.  There are several choices for how to do this, ranging
    from Suzuki-Trotter decompositions that are most natural for Hamiltonians
    with short-range interactions~\cite{vidal2004efficient}, to time-step
    targeting and other
    techniques~\cite{feiguin2005time,zaletel2015time,ronca2017time} designed for
    Hamiltonians with long-range interactions.
    
    We use the classical 4th-order Runge-Kutta (RK4) algorithm together with an
    MPO representation of the full Hamiltonian to carry out a single time-step
    propagation,
    \begin{align}
        | k_1 \rangle & = -i \hat{H}(t) | \Psi(t) \rangle  \nonumber \\
        | k_2 \rangle & = -i \hat{H}(t + \tau/2) 
            (| \Psi(t) \rangle + \frac{1}{2}
            \tau | k_1 \rangle) \nonumber \\
        | k_3 \rangle & = -i \hat{H}(t + \tau/2) 
            (| \Psi(t) \rangle + \frac{1}{2} 
            \tau | k_2 \rangle) \nonumber \\
        | k_4 \rangle & = -i \hat{H}(t + \tau) 
            (| \Psi(t) \rangle +
            \tau | k_3 \rangle) \nonumber \\
        | \Psi(t+\tau) \rangle & = | \Psi(t) \rangle + \frac{1}{6} \tau(| k_1 \rangle +
        2 | k_2 \rangle + 2| k_3 \rangle + | k_4 \rangle)  \label{eq:RK4}
    \end{align}
    Here, $\tau$ is the time step, $\hat{H}(t)$ is an MPO and each wavefunction
    $|k_i\rangle$ is an MPS.  The stable region for the time step $\tau$
    requires the RK4 propagator to have modulus $< 1$. As described in the
    Supporting Information (SI) (section 1), for real time propagation, $\tau
    \cdot e < 2.828$, and for imaginary time propagation, $\tau \cdot e <
    2.785$, where $e$ is the absolute maximal eigenvalue of $\hat{H}$.  In
    practice, to simulate with as large a time-step as possible, one usually
    defines $\hat{H}(t)$ with the lowest state energy subtracted. Note that for
    time independent Hamiltonians, the RK4  method reverts back to eq.
    \eqref{eq:taylerexpansion}, which is nothing but a 4th order Taylor
    expansion of $e^{-i\hat{H}\tau}$ around time $0$.
    \begin{gather}
        e^{-iH\tau}   
        \approx 1 + (-iH\tau) + \frac{(-iH\tau)^2}{2!} +
        \frac{(-iH\tau)^3}{3!} + \frac{(-iH\tau)^4}{4!}  \label{eq:taylerexpansion}
    \end{gather}
    
\begin{figure*}[htbp]
    \centering
    \includegraphics[width = 1.0 \textwidth]{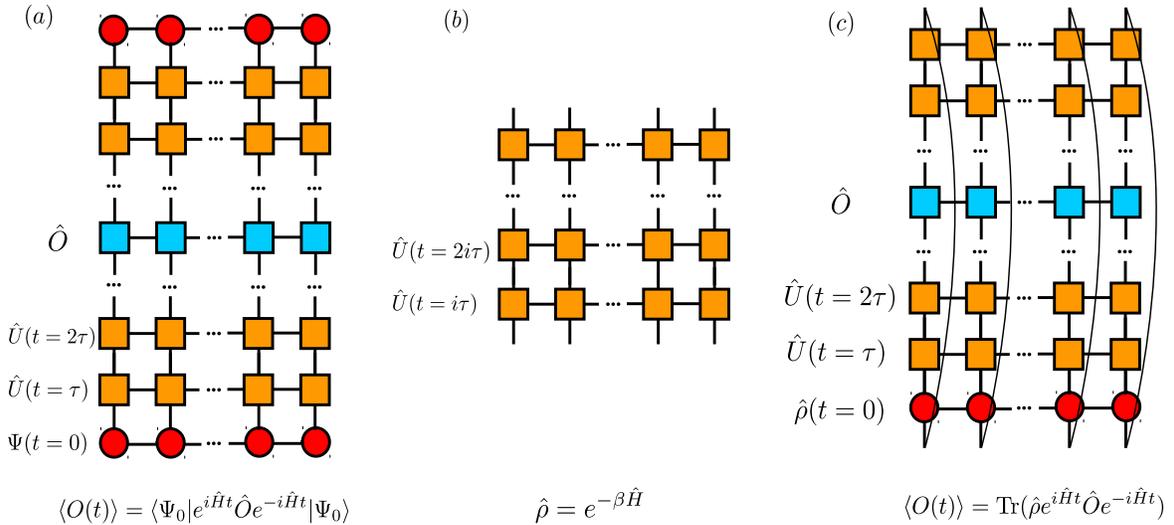}

    \caption{\label{fig:operation}
        The graphical representation of (a) $\langle \hat{O}(t)\rangle$ at zero
        temperature (b) the density operator $e^{-\beta \hat{H}}$ (c)  $\langle
        \hat{O}(t) \rangle$ at finite temperature. The long line linking the top
        and bottom physical bonds represents a trace.  Compression is carried
        out after each application of the unitary propagator in ascending time
        in (a) ,(b) and (c).
        }
    \end{figure*}

    During each time step propagation, $\hat{H} \cdot \Psi$ is carried out for 4
    times. At each time the virtual bond dimension $a_n$ of MPS is enlarged by a
    factor of $D$, and thus needs to be compressed.  We carry out compression
    back to the desired bond dimension $M$ using the SVD algorithm described
    previously.  Overall, the graphical representation of the expectation value
    $\langle O(t) \rangle$ at zero temperature computed using ZT-TD-DMRG is
    shown in Figure \ref{fig:operation}(a).
    
    For finite temperature TD-DMRG (FT-TD-DMRG), we must first represent the
    initial thermal state in matrix product form.  The most common method to
    achieve this is via the thermo field dynamics approach (or called ``purification'' or ``ancilla'' approach).
    \cite{takahashi1975takahashi,verstraete2004matrix,zwolak2004mixed,
    feiguin2005finite, borrelli2016quantum} The basic idea is to represent a
    mixed state density operator as a partial trace of a pure state density
    operator in an enlarged space, namely, the physical space $P$ $\otimes$ an
    auxiliary space $Q$. The simplest choice is to choose $Q$ identical to the
    physical space $P$,
    \begin{gather}
        \hat{\rho} = \sum_i s_i | i \rangle \langle i | = \textrm{Tr}_Q  | \Psi
        \rangle \langle \Psi | \nonumber \\ 
        | \Psi \rangle =  \sum_i s_i^{1/2} | i \rangle_P | i \rangle_Q 
    \end{gather}
    Then, the thermal equilibrium density operator can be expressed as
    \begin{gather}
        \hat{\rho}_{\textrm{eq}} = \frac{e^{-\beta \hat{H}}}{Z(\beta)} = 
        \frac{\textrm{Tr}_Q |\Psi_\beta \rangle \langle \Psi_\beta |} 
        {\textrm{Tr}_{PQ} |\Psi_\beta \rangle \langle \Psi_\beta |}  \nonumber \\
        | \Psi_\beta \rangle = e^{-\beta \hat{H}/2} | \Psi_\infty \rangle, \, 
        | \Psi_\infty \rangle = \sum_i | i \rangle_P | i \rangle_Q  
    \end{gather}
    where $\hat{H} = \hat{H}_P \otimes \hat{I}_Q$ and $| \Psi_\infty \rangle$ is
    an unnormalized product of maximally entangled states between each physical
    site and a corresponding auxiliary site.  To obtain $| \Psi_\beta \rangle$,
    imaginary time propagation is carried out on the initial state $|
    \Psi_\infty \rangle$ and once temperature $\beta$ is reached, real time
    propagation is carried out on $| \Psi_\beta \rangle$. Since all operations
    are on pure states, the finite temperature algorithm can be implemented
    using exactly the same code as the zero temperature algorithm, the only
    difference being that the total size of the system (i.e. the number of
    sites) is increased by a factor of 2.  However, in terms of MPO's, we can
    equivalently understand the finite temperature method as a direct
    propagation of the density matrix, eliminating the need for
    ancillas~\cite{barthel2013precise} (see Figures \ref{fig:operation}(b) and
    (c)).  In this case, the initial infinite temperature density matrix
    $|\Psi_\infty \rangle$ is simply an identity operator in the Hilbert space,
    and is an MPO with bond dimension 1. Then the thermal equilibrium density
    operator can be expressed as 
    \begin{gather}
        e^{-\beta \hat{H}} = e^{-\beta \hat{H}/2} \hat{I} \hat{I} e^{-\beta
        \hat{H}/2} 
    \end{gather}
    To carry out the MPO time propagation, MPO compression is carried out at
    each step by interpreting an MPO as an MPS where the two physical indices on
    each site are viewed as the physical index of an MPS in an enlarged space,
    making the MPO compression identical to the MPS compression in the ancilla
    picture. Note that the initial high-temperature density matrix $\hat{I}$ can
    in principle be replaced by any unitary operator $\hat{U}$, since
    \begin{gather}
        e^{-\beta \hat{H}} = e^{-\beta \hat{H}/2} \hat{U} \hat{U}^\dagger e^{-\beta
        \hat{H}/2} 
    \end{gather}
    and this degree of freedom has recently been used to reduce the entanglement
    growth and thus cost of time propagation.\cite{karrasch2012finite} In our
    FT-TD-DMRG implementation, we use the MPO propagation formulation starting
    from an initial identity operator. 

    The most time-consuming part in the TD-DMRG algorithm is the SVD
    compression. The computational scaling of the SVD compression is roughly
    $O(Nk_e k_\nu pM^3D^3)$ for ZT-TD-DMRG and $O(N k_e k_\nu p^2M^3D^3)$ for
    FT-TD-DMRG ($N$ is the number of propagation steps). We see that the only
    difference in formal scaling arises from the increase in physical bond
    dimension going from ZT to FT ($p$ to $p^2$). However, the bond dimension
    $M$ necessary for a given accuracy will be different in the finite
    temperature and zero temperature formulations.
    To see this, consider the zero temperature limit or a pure state, where the density matrix in the FT-TD-DMRG formulation
    can be trivially expressed as
    \begin{align}
        \rho = &| \Psi \rangle \langle \Psi | \nonumber \\ 
		= &	\sum_{\{a,a'\},\{\sigma, \sigma'\}}
       (A^{\sigma_1}_{a_1} A^{\sigma'_1}_{a'_1}) (A^{\sigma_2}_{a_1a_2} A^{\sigma'_2}_{a'_1a'_2}) \cdots
       (A^{\sigma_n}_{a_{n-1}}A^{\sigma'_n}_{a'_{n-1}})  \nonumber \\ 
         & \qquad \qquad | \sigma_1\sigma_2\cdots\sigma_n \rangle  \langle  \sigma_1'\sigma'_2\cdots\sigma'_n | \nonumber \\
		= &	\sum_{\{b\},\{\sigma, \sigma'\}}
        B^{\sigma_1, \sigma'_1}_{b_1} B^{\sigma_2, \sigma'_2}_{b_1 b_2} \cdots 
		B^{\sigma_n, \sigma'_n}_{b_{n-1}}| \sigma_1\sigma_2\cdots\sigma_n \rangle  \langle  \sigma_1'\sigma'_2\cdots\sigma'_n |
	\end{align}
    where $\textrm{dim}(b_i) =  \textrm{dim}(a_i) \cdot \textrm{dim}(a'_i)$. This shows that the bond dimension for a given accuracy in the
    FT-TD-DMRG at low temperature is the square of that in the ZT-TD-DMRG formulation.
    
    With ZT-TD-DMRG and FT-TD-DMRG, the absorption and fluorescence spectra can be
    calculated by taking the Fourier transform of the dipole-dipole
    time correlation function.
    \begin{gather}
        \sigma_{\textrm{abs}(\textrm{emi})}(\omega) \propto \frac{1}{2\pi}
        \int_{-\infty}^{+\infty}
        dt e^{i\omega t} C(t) \label{eq:spectrum} \\
		C(t) = \langle \hat{\mu}(t) \hat{\mu}(0) \rangle_{g(e)}
    \end{gather}
    The subscript $e,g$ represents the one exciton excited state space and zero
    exciton ground state space respectively. The bra-ket denotes the lowest
    energy state expectation value at zero temperature and the thermal
    equilibrium average at finite temperature.  The frequency dependence of the
    prefactor is respectively $\omega$ and $\omega^3$ for the absorption and
    emission cross section.
    To apply a Gaussian broadening in the frequency domain (convolving the original spectrum with a Gaussian function), we
    multiply the dipole-dipole correlation function in the time domain with a Gaussian function (pointwise) before the Fourier transform
    \begin{gather}
        C'(t) =  C(t) e^{-(t/t_{\textrm{Gaussian}})^2}
    \end{gather}
    where the connection between broadening in the time domain and in the frequency domain is provided by the Convolution Theorem.

\subsection{Computational optimizations} \label{sec:techniques}
    We now consider some techniques to improve the efficiency and accuracy of
    time propagation for the specific case of the Holstein Hamiltonian.  When
    calculating the linear spectrum, the excited state is in the one-exciton
    space, while the ground state in the zero-exciton space. In the zero-exciton
    space, the Hamiltonian in eq. \eqref{eq:discreteH} becomes trivial --- only
    the vibrational energy terms survive. These terms are local and commute with
    each other,
    \begin{gather}
        \hat{H}_g = \sum_{in} \omega_{in} b_{in}^\dagger b_{in}
    \end{gather}
    Thus, unlike the propagator in the one-exciton space that is approximated by
    the RK method, the propagator in the zero-exciton space can be exactly
    represented as an MPO with bond dimension 1,
    \begin{gather}
        e^{-i\hat{H_g}t} = e^{-i\hat{H}_{g1}t} e^{-i\hat{H}_{g2}t} \cdots
        e^{-i\hat{H}_{gn} t}
    \end{gather}
    Therefore, time propagation in the zero-exciton space is exact and MPS
    compression is not required.
    
    For example, the dipole-dipole time correlation function in
    eq. \eqref{eq:spectrum} for the zero temperature fluorescence spectrum is 
    \begin{gather}
        C(t) = e^{iE_{e}t}\langle \Psi_e(0) | \hat{\mu} e^{-i\hat{H}_gt}
        \hat{\mu} | \Psi_e(0) \rangle
    \end{gather}
    Here, $| \Psi_e(0) \rangle$ and $E_e$ are the wavefunction and energy of the
    lowest excited state in the one-exciton space. Since the single lowest
    excited state can be accurately obtained with the standard DMRG algorithm
    with a modest virtual bond dimension, and the subsequent time propagation
    under $\hat{H}_g$ does not increase the bond dimension, the zero temperature
    fluorescence spectrum can be calculated without the cost of a full TD-DMRG
    calculation. Note that the same technique can be applied to other
    observables where propagation is restricted to the zero-exciton space.
    
    Another potential optimization in TD-DMRG is to carry out a basis
    transformation to minimize the growth of entanglement. In the Holstein
    Hamiltonian in eq. \eqref{eq:discreteH} every exciton site is coupled to a set
    of vibration sites, leading to a ``star-like'' topology for the exciton-phonon
    interactions.  An alternative topology is a ``chain'' topology that can be
    obtained via a unitary rotation of the vibrational
    basis.~\cite{de2015discretize, wolf2014solving} We have compared the two representations in
    our calculations and found that the ``star'' representation generally
    introduces less entanglement (see SI section 3). Thus we use the ``star''
    representation in our calculations.

\section{Results}
\label{sec:results}
\subsection{Linear spectrum of PBI J-type aggregates}
    From a purely excitonic coupling perspective, two typical aggregation types
    have been defined by Kasha.\cite{kasha1963energy} One is the J-type aggregate,
    where the sign of the excitonic coupling is negative (assuming the transition
    dipole moments all point in the same direction), as is the case for molecules packed in
    a ``head-to-tail'' orientation. In the lowest (highest) excitonic band, the
    dipole moments of the molecules interfere constructively (destructively), so
    the absorption and fluorescence are enhanced and red shifted. Another is the
    H-type aggregate, in which the molecules are stacked in a ``side-by-side''
    orientation, resulting in a positive excitonic coupling. In contrast to
    the J-type aggregates, the absorption is blue shifted and the fluorescence is
    suppressed.
    
    Perylene bisimide dyes (PBI) are prototypical building blocks for H- and J-
    type molecular aggregates, and are potential candidates for an artificial
    light harvesting system.\cite{wurthner2004perylene} The optical properties
    have been investigated both experimentally and theoretically.  Among the
    reported theoretical calculations of the linear spectrum of J-type PBI, the
    ML-MCTDH calculations by K\"{u}hn \textit{et al} are probably the most
    accurate. The first set of ML-MCTDH calculations were presented in
    Ref.~\citenum{ambrosek2012quantum} with the largest system treated being a
    linear hexamer, including up to 5 vibrational modes per molecule.
    Subsequently, improved calculations including 10 modes per molecule were
    reported.  \cite{schroter2015exciton} The ML-MCTDH calculations were all at
    zero temperature and the finite temperature spectrum was not computed.

    To verify the correctness of our TD-DMRG implementation, we first calculated
    the zero temperature linear absorption of a PBI chain, and compared our
    results to those from ML-MCTDH, using the parameters of the ML-MCTDH
    calculations with 10 phonon modes.\cite{schroter2015exciton} We used a total
    evolution time $\tau$ of 20 a.u, $N=20000$ time steps and an SVD cutoff of
    $10^{-3}$. The maximal virtual bond dimension is 42.  We applied a Gaussian
    broadening $C(t)e^{-(t/t_{\textrm{Gaussian}})^2}$ in the time domain with
    $t_{\textrm{Gaussian}}=2000 \, \textrm{fs}$.

\begin{figure}[htbp]
    \centering
    \includegraphics[width = 0.48 \textwidth]{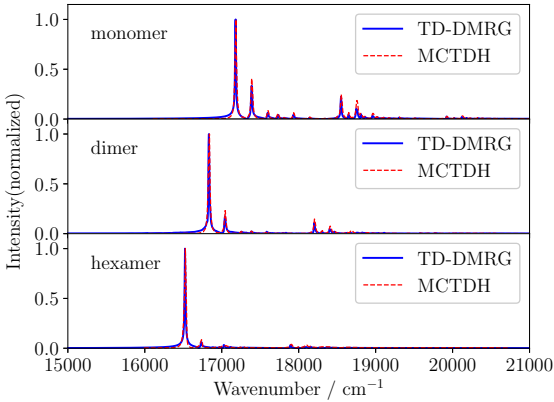}

    \caption{\label{fig:PBI}
        The calculated zero temperature linear absorption spectra of the PBI
        monomer, dimer and hexamer from ZT-TD-DMRG. Each molecule includes 10
        normal modes. The ML-MCTDH results are from
        Ref.~\citenum{schroter2015exciton}
        }
    \end{figure}

    In Figure \ref{fig:PBI}, we show the zero temperature absorption spectrum
    computed from ZT-TD-DMRG. The spectrum is essentially identical to that
    obtained by ML-MCTDH. Two main vibronic structures can be identified. One is
    due to a high frequency mode $\omega=1371 \textrm{cm}^{-1}$, $S=0.208$ and
    the other is due to two modes with similar frequencies and Huang-Rhys
    factors $\omega=206, 211 \textrm{cm}^{-1}$, $S=0.197, 0.215$. The remaining
    modes cannot be clearly assigned because of their small Huang-Rhys factors
    (<0.1).
    
\begin{figure}[htbp]
    \centering
    \includegraphics[width = 0.48 \textwidth]{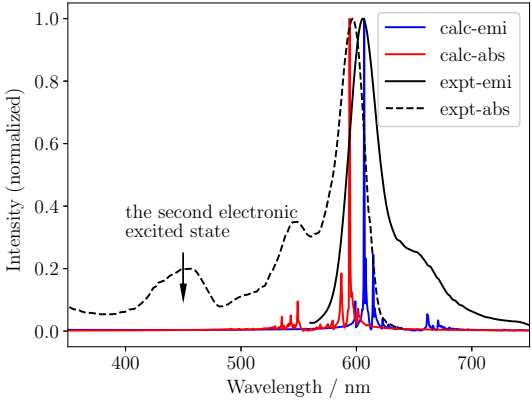}

    \caption{\label{fig:FTPBI}
        The calculated absorption and fluorescence spectra of PBI dimer at 298K
        from FT-TD-DMRG. The SVD cutoff is $10^{-3}$, total simulation time
        $\tau=20 \textrm{a.u}$, and total number of time-steps  $N=5000$.  The
        maximal virtual bond dimension is 120.  No broadening is applied. The
        experimental spectrum from Ref.~\citenum{li2008highly} is also plotted.
        }
    \end{figure}

    Using our FT-TD-DMRG implementation, we can further extend our calculations
    to obtain the finite temperature absorption and fluorescence spectra. We
    calculated the spectrum of the PBI dimer at 298K which can be compared
    against the reported experimental spectrum \cite{li2008highly} (see Figure
    \ref{fig:FTPBI}). We see that the spectra agree well but the 0-0 transition
    peak is slightly shifted from the experimental position. The fine structure
    of the room temperature spectrum is also well reproduced, except for the
    absorption peak near 450nm. This peak originates from the second local
    electronic excited state,\cite{ambrosek2012quantum}  which is not contained
    in our model Hamiltonian. 
    Unfortunately, there are no finite temperature MCTDH results reported for this system for comparison.

    In conjunction, our calculations on the PBI systems show that TD-DMRG is a
    high level quantum dynamics method and a practical alternative to ML-MCTDH,
    achieving good accuracy at finite as well as at zero temperatures.
    
\subsection{TD-DMRG vs. the n-particle approximation}
    The n-particle approximation is a popular method to calculate the linear
    spectrum of molecular aggregates. However, for small n, it is only accurate
    in the weak excitonic coupling regime. We now compare TD-DMRG with 1-, 2-,
    3- particle approximations in a model chain system, composed of 4 identical
    molecules, with nearest neighbor excitonic coupling and periodic boundary
    conditions. We choose each molecule to have a single vibration mode and for
    each mode to have up to 8 phonons. Full ED can be carried out as an exact
    reference in this system.  We use the n-particle approximation with respect
    to the Hamiltonian and basis in eq. \eqref{eq:discreteH}, without first
    carrying out a Lang-Firsov transformation as is done in some other
    works.\cite{philpott1967vibronic, hoffmann2002optical, spano2009spectral,
    wenqiang2016effect} Note that the Lang-Firsov transformation for an infinite
    phonon space is unitary, but for a truncated phonon space it is not. Thus
    the results with or without Lang-Firsov transformation will be a little bit
    different, but this difference will be negligible at low temperatures.

\begin{figure*}[htbp]
    \centering
    \includegraphics[width = 0.8 \textwidth]{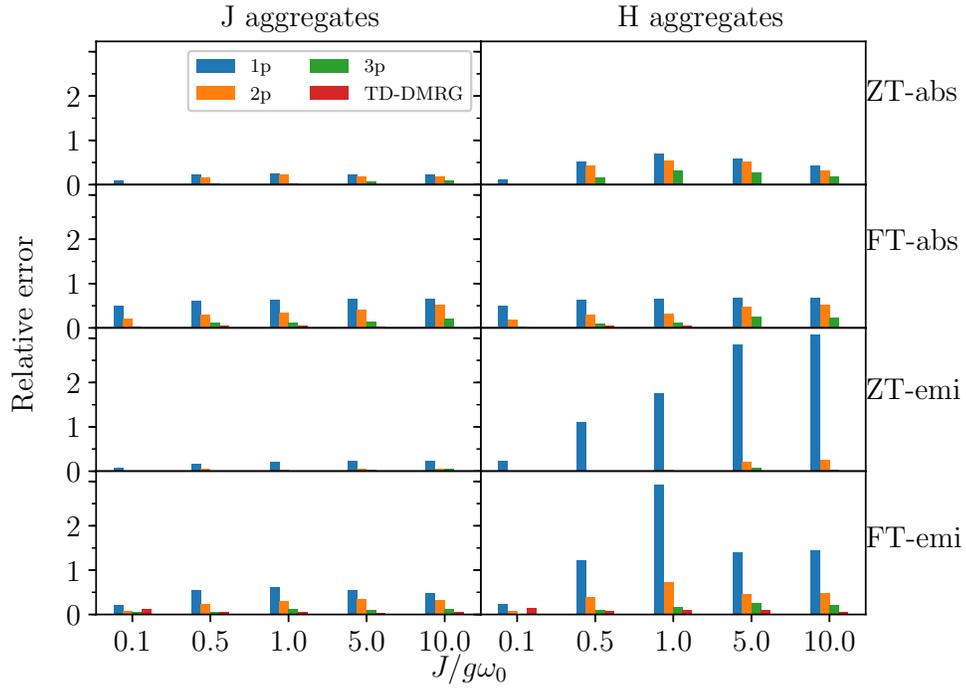}

    \caption{\label{fig:vsNp}
         The relative error of the absorption and emission spectra at both zero
         and finite temperature from four different methods, TD-DMRG (red),
         1-(blue), 2-(orange), 3-(green) particle approximation methods.  The
         excitonic / exciton-phonon coupling ratio is 0.1, 0.5, 1.0, 5.0,
         10.0.
		}
    \end{figure*}

    We calculated the zero and finite temperature absorption and fluorescence
    spectra for both J- and H- type aggregates.  We set the Huang-Rhys factor to
    1.0, the temperature to be $k_b T = \hbar \omega_0$ and varied the excitonic
    coupling to the exciton-phonon coupling ratio $J/g \omega_0$.  The SVD
    cutoff in ZT-TD-DMRG was set to $10^{-4}$ and that of FT-TD-DMRG to
    $10^{-3}$.  For a direct comparison with TD-DMRG, the full ED and n-particle
    approximation spectra were also obtained by time-propagation. All
    propagations were carried out with a time step
    $\tau=\frac{0.032}{\omega_0}$ and a total number of steps $N=20000$. A small
    Gaussian broadening was applied to increase the smoothness of the spectrum,
    $t_{\textrm{Gaussian}} = \frac{131}{\omega_0}$.  To compare the various
    spectra, we use the relative error of the spectrum compared to the exact
    result from full ED,
    \begin{gather}
        \textrm{relative error}_{\textrm{method}} = 
        \frac{\sum_{i=1}^{N}|\sigma_{\textrm{method}}(\omega_i) -
        \sigma_{\textrm{exact}}(\omega_i)|d\omega}{
            \sum_{i=1}^{N}\sigma_{\textrm{exact}}(\omega_i) d\omega}
    \end{gather}
    where $\omega_i$ is a discrete point in the frequency domain and
    $\sigma(\omega_i)$ is the strength at frequency $\omega_i$. The relative
    error of the different methods is shown in Figure \ref{fig:vsNp}. Note that
    in the n-particle approximation, since even the 0-0 energy gaps changes with
    n, we shift all the spectra by subtracting the 0-0 energy gap.
	
    Our calculations show that the accuracy of TD-DMRG greatly surpasses that of
    n-particle approximation methods in all cases.  The error of FT-TD-DMRG is a
    little larger than that of ZT-TD-DMRG calculation, probably due to the
    larger cutoff threshold in the SVD. In the coupling regime where $J$ and
    $g\omega_0$ are comparable, also known as the strongly correlated regime,
    the virtual bond dimension of the MPS increases much more rapidly than in the
    two limiting coupling regimes. This rapid increase in bond dimension
    coincides with the breakdown of perturbation theory in this regime.

\begin{figure}[htbp]
    \centering
    \includegraphics[width = 0.45 \textwidth]{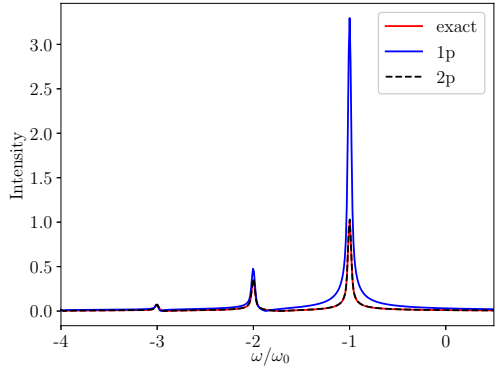}

    \caption{\label{fig:1pemi}
        The zero temperature emission of a H-type chain model for $J/g \omega_0 =
        1$. The spectrum is scaled by a factor which normalizes the exact
        spectrum. The 0-0 position is shifted to the origin. The results
        of three methods are shown, exact(red), 1-particle(blue), 2-particle(black dotted)
        approximation.
        }
    \end{figure}

    For the n-particle approximation, we find a wide variation in the accuracy
    depending on the simulation regime. In particular, we find that 
	\begin{enumerate*}[label=(\roman*)]
        \item In the weak excitonic coupling regime $J/g \omega_0=0.1$, the
            n-particle approximations perform well. However, when $J/g \omega_0$
            is larger than 0.5, the error is large. There is no simple trend in
            the error as a function of $J/ g\omega_0$.
        \item The finite temperature spectrum is worse than the zero temperature
            spectrum. When the temperature is increased, in addition to the
            single electronically excited molecule itself, additional molecules in the
            ground state become vibrationally excited, which are not included in
            the n-particle approximation. Therefore, the n-particle approximation is
            only good for targeting the lowest energy states. When high
            energy states contribute to the spectrum, the results worsen.
        \item The 1-particle approximation results for the emission of H-type
            aggregates are not reliable, as they overestimate the 0-1
            emission (see Figure \ref{fig:1pemi}). Taking the dimer as an example, the
            vector $|e_1 \nu_1, e_2 \nu_2 \rangle$ denotes an occupation
            representation of the dimer basis. ($|e_i \rangle$ represents an exciton,
            $|\nu_i \rangle$ represents a vibrational mode). In the 1-particle approximation space,
            only the $|10,00 \rangle$, $|00,10 \rangle$, $|11,00\rangle$,
            $|00,11\rangle$ states are considered. The last two basis states contribute to the 0-1
            emission strength.  The $|01,10\rangle$, $|10,01\rangle$ states
            which are included in the 2-particle approximation, do not appear.
            However, these two neglected basis states are important because they
            directly electronically couple to $|11,00\rangle$, $|00,11\rangle$. In
            H-type aggregates, the wavefunction amplitudes for $|11,00\rangle$ and
            $|01,10\rangle$ (or $|10,01\rangle$ and $|00,11\rangle$) are
            opposite in sign, so they contribute to decreasing the 0-1 strength
            from that when only the $|11,00\rangle$, $|00,11\rangle$ states are
            in the Hilbert space, as in the 1-particle approximation.  This
            explains why the 2-particle approximation improves the 0-1 emission
            dramatically.  We thus recommend that the 2-particle approximation
            is the lowest level n-particle approximation to use when calculating
            the fluorescence spectrum of H-type aggregates.
	\end{enumerate*}

\subsection{TD-DMRG: sources of error}
    To better understand  the errors in the TD-DMRG propagation, we can separate the two sources of error that arise, namely, from the DMRG
    compression and from the RK evolution.  To study the DMRG error by itself (at zero temperature), at each time step $t \to t+\tau$, we
    expand the wavefunction $\Psi_{\textrm{DMRG}}(t)$ in the DMRG representation into a full configuration interaction (FCI) representation
    $\Psi_{\textrm{FCI}}(t)$, which is then propagated by an exact propagator (calculated by ED) to $\Psi_{\textrm{FCI}}(t+\tau)$, before
    being compressed into the DMRG representation $\Psi_{\textrm{DMRG}}(t+\tau)$ by SVD under a specified compression criterion. To analyze
    the RK error by itself, all steps are performed within the FCI representation.

\begin{figure}[htbp]
        \centering

        \subfloat[]{
            \label{fig:error}
            \includegraphics[width = 0.45 \textwidth]{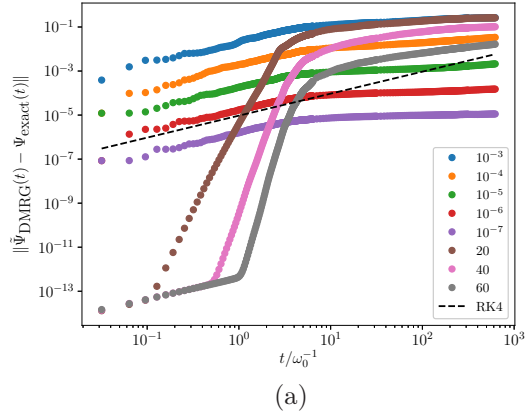}  
        } \\
        \subfloat[]{
            \label{fig:M}
            \includegraphics[width = 0.45 \textwidth]{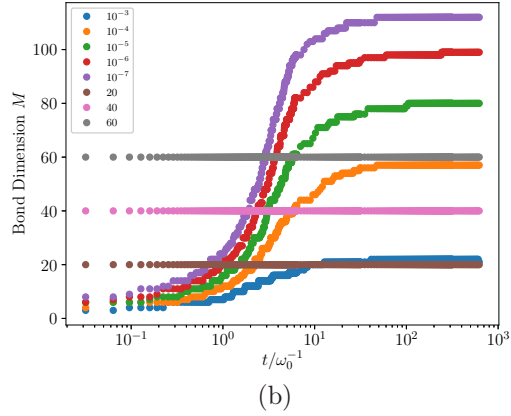}
        }

        \caption{
            (a) The error $\|\Psi_{\textrm{DMRG}}(t)-\Psi_{\textrm{exact}}(t)\|$ and 
          (b) the maximal bond dimension $M$ of the DMRG wavefunction as a function of  time under different compression
          criteria (fixed SVD cutoffs: $10^{-3}$,
          $10^{-4}$, $10^{-5}$, $10^{-6}$, $10^{-7}$ and fixed bond dimensions: 20, 40 and 60).  The error due to the RK4 integration is also plotted (black
            dashed line).
             }
    \end{figure}

    In Figure \ref{fig:error}, we plot the error of the TD-DMRG wavefunction ($\Psi(t) = e^{-i\hat{H}t} \hat{\mu} \Psi_{\textrm{g}}(0)$) due
    to  the DMRG compression alone for different SVD cutoffs or bond dimensions $M$, for the same model J-type aggregates with $J/g\omega_0
    = 1.0$ (the challenging intermediate coupling regime). The time step $\tau$ and the total number of propagation steps $N$ are the same
    as in the previous section.
    For fixed SVD cutoff, the wavefunction error initially grows smoothly, with a polynomial growth as a function of propagation time.
    Interestingly, for fixed  bond dimension, the error versus time shows a ``three-stage'' structure. The error is initially quite small,
    and then grows very quickly, and then finally reaches a regime where it increases smoothly with time similarly to the case of fixed  SVD
    cutoff. In Figure \ref{fig:M}, the bond dimension versus time is also plotted. When the SVD cutoff is fixed, the maximal bond dimension grows up
    to a certain value instead of growing indefinitely, and this value increases as the SVD cutoff is tightened. Thus, comparing Figures \ref{fig:error} and \ref{fig:M}, the ``three-stage''
    structure of the error for fixed bond dimension $M$ arises because
    \begin{enumerate*}[label=(\roman*)]
    \item in the first stage, $M$ is larger than the required $M_{\textrm{req}}$ for exact evolution, giving a very small total error
        (controlled by the round-off error);
    \item in the second ``rapid growth'' stage, the $M_{\textrm{req}}$ for maintaining a given accuracy increases very quickly past the
        fixed $M$, thus the error increases rapidly;
    \item finally in the third stage, the growth rate of $M_{\textrm{req}}$ for a certain accuracy slows down, thus the error grows smoothly again.
    \end{enumerate*}

    The formal relation between the global error due to the RK4 integration and  time (total time $t$ and time step $\tau$) is well-known
    and is $O(t\tau^4)$.  In our problem, the error due to RK4 integration is shown in Figure S2, which is consistent with the above
    relation.  
    
\begin{figure}[htbp]
    \centering
    \includegraphics[width = 0.45 \textwidth]{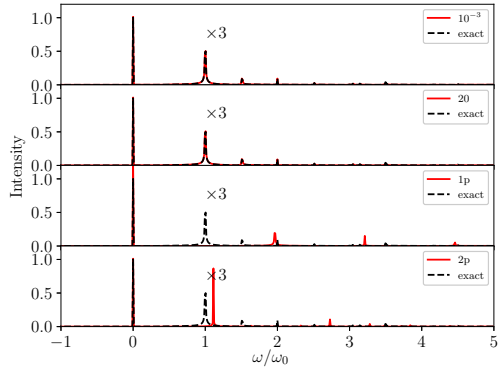}

    \caption{\label{fig:comp}
        The absorption spectrum at zero temperature calculated by TD-DMRG (SVD cutoff $= 10^{-3}$), TD-DMRG ($M=20$), 1-particle
        approximation and 2-particle approximation. The exact result is also plotted for comparison. (To show the fine structure, the
        absorption strength is multiplied by 3
        when $\omega/\omega_0 > 0.5$).
        }
    \end{figure}

    Comparing the RK4 integration error and the DMRG compression error within the time window calculated, the error due to RK4 is smaller
    than that due to DMRG compression, for the SVD cutoffs of $10^{-3}$ and $10^{-4}$ (see Figure \ref{fig:error}, black dashed line).  In
    practice, an SVD cutoff of $10^{-3}$ or $10^{-4}$ is usually a good balance between accuracy and efficiency, thus we expect that the
    DMRG compression error will typically be the main source of error rather than the RK4 error. However, note that even though the DMRG
    truncation for bond dimension $M=20$ or SVD cutoff $=$ $10^{-3}$ leads to a seemingly large wavefunction error of approximately $10\%$,
    the obtained spectrum is still     very accurate (see Figure \ref{fig:comp}), except for some small shifts in the high frequency region
    where $\omega / \omega_0 > 2$.  This is quite different from the 1-particle and 2-particle approximations, where all the 0-n transitions
    amplitudes are calculated with large errors except for the 0-0 transition.

    \subsection{Distyrylbenzene H-type aggregates}

\begin{figure}[htbp]
    \centering
    \subfloat[]{
        \includegraphics[width = 0.065 \textwidth]{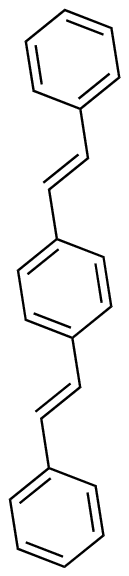}
    } \qquad
    \subfloat[]{
        \includegraphics[width = 0.25 \textwidth]{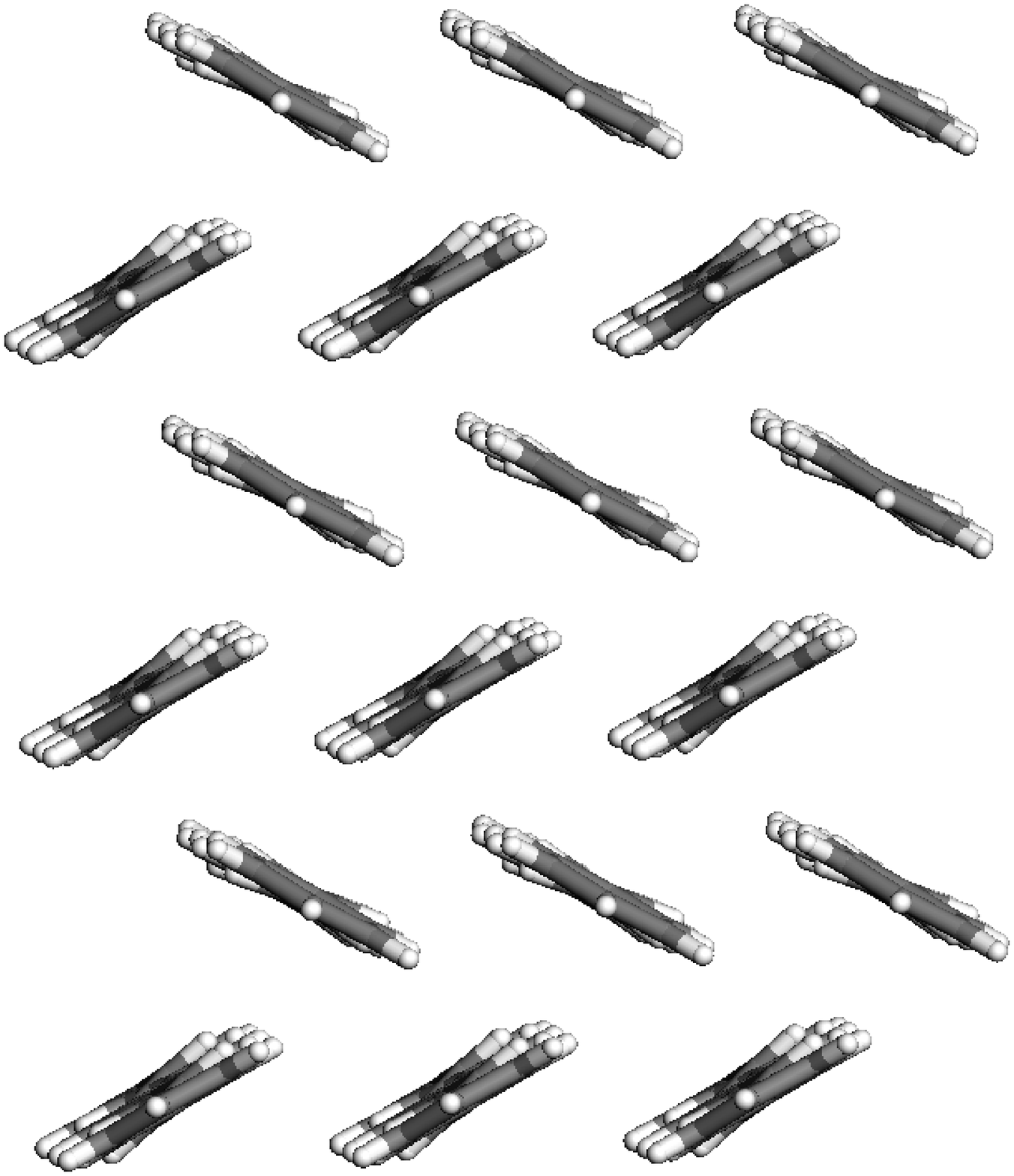}
    } \\
    \subfloat[]{
        \includegraphics[width = 0.48 \textwidth]{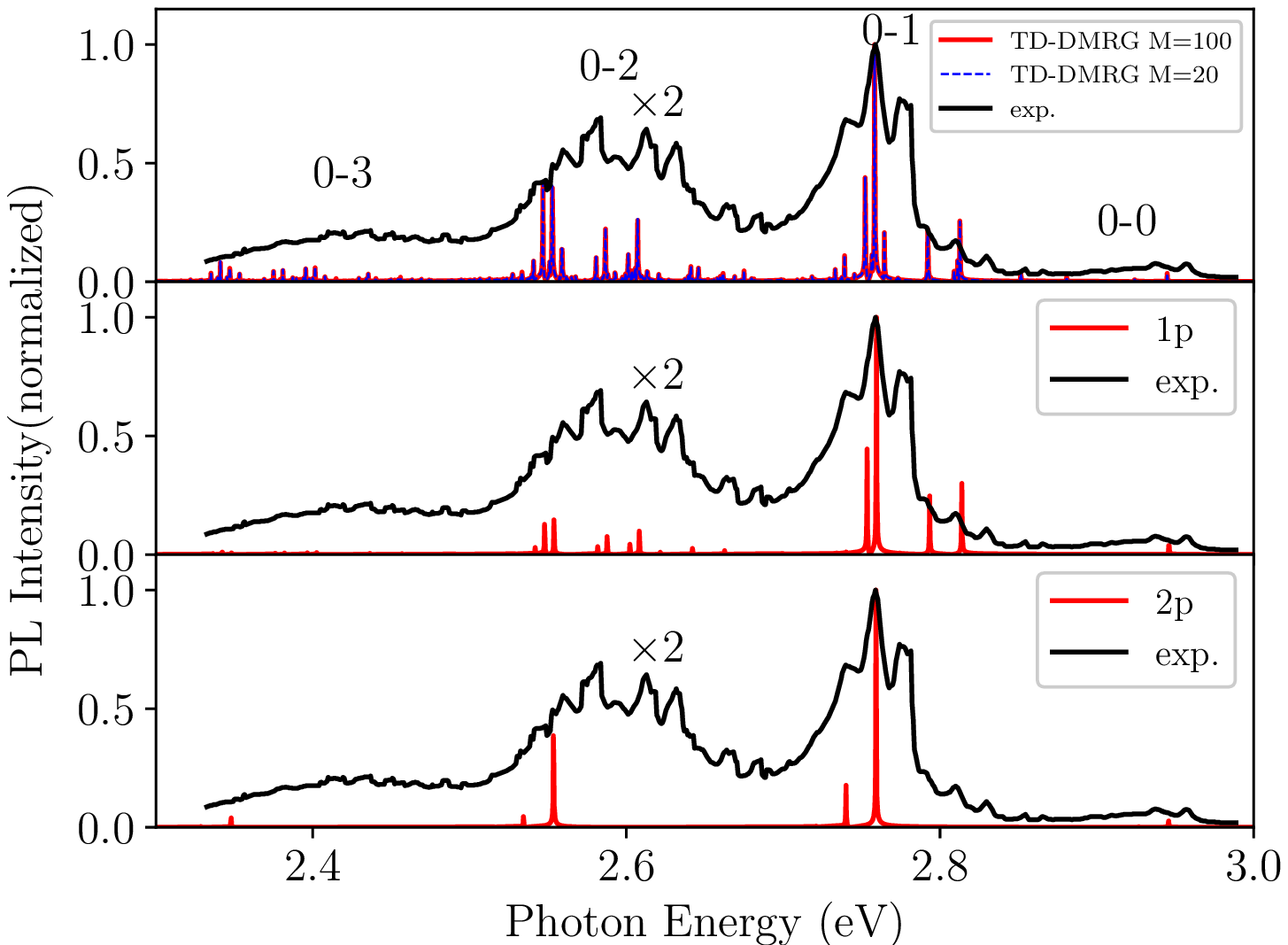}
    }

    \caption{\label{fig:DSB}
        (a) The chemical structure of the DSB molecule.
        (b) The packing structure of the selected DSB aggregates with 18
        molecules.
        (c) The zero temperature fluorescence spectrum of DSB aggregates from
        TD-DMRG (red: $M=100$, dashed blue: $M=20$), and the 1-, 2- particle approximations. $\tau=20 \textrm{a.u}, N=30000$, no
        broadening is applied. The peaks below 2.7 eV are multiplied by 2 to
        show the fine structure. The experimental results at 1.4K from
        Ref.~\citenum{wu2003optical} are also plotted (black line). 
        }
    \end{figure}

    We conclude our study by considering a TD-DMRG calculation in a realistic
    H-type aggregate consisting of distyrylbenzenes (DSB).\cite{wu2003structual}
    In the DSB crystal, the DSB molecules pack into an intralayer side-by-side
    herringbone structure shown in Figure \ref{fig:DSB}(b). The excitonic
    coupling between layers can be neglected. Along the direction perpendicular
    to the layer, the transition dipole moments of the intralayer molecules all
    align in parallel and form an H-type structure. In addition, there is a
    small component of the transition dipole moment forming a J-type structure
    parallel to the layer, resulting in an anisotropic effect of the aggregation
    on the spectra. We only consider the dominant H-type structure here. We
    choose a cluster with 18 molecules as our system (see Figure
    \ref{fig:DSB}(b)). The parameters of the excitonic coupling and
    exciton-phonon coupling are adopted from Ref.~\citenum{wenqiang2016effect}
    (see SI section 4). The linear spectrum of the system has been investigated
    at a qualitative level using n-particle approximation methods within the
    Holstein exciton model and the general features of the spectra have been
    obtained, but the fine structure has not been reproduced because only a few
    ($<5$) effective modes have been considered.\cite{spano2001absorption,
    spano2003fundamental, zhao2005vibronic, spano2006excitons} For our TD-DMRG
    calculations, we chose 14 normal modes for each molecule with a Huang-Rhys
    factor greater than 0.02. To our knowledge, this is the largest DSB molecular
    aggregate so far studied for which the fluorescence spectrum is calculated
    non-perturbatively.  The corresponding Hilbert space size for the TD-DMRG
    (as well as various n-particle approximations) is listed in
    Table.\ref{table:DSBtable}.

    \begin{table*}[bht]
    \centering
    \caption{System sizes for TD-DMRG and 1-, 2- particle approximation methods}
    \label{table:DSBtable}
    \begin{tabular}{ *5c }
        \hline
         &  N molecules & N modes & N phonons & dimension of Hilbert space \\ 
        \hline
         TD-DMRG & 18 & 14 & 10 & $18 \times 10^{252}$ \\
         1-particle    & 18 & 5  & 4  & 18432 \\
         2-particle    & 18 & 2  & 4  & 73728 \\
        \hline
    \end{tabular}
	\end{table*}
    
\begin{figure}[htbp]
    \centering
    \includegraphics[width = 0.45 \textwidth]{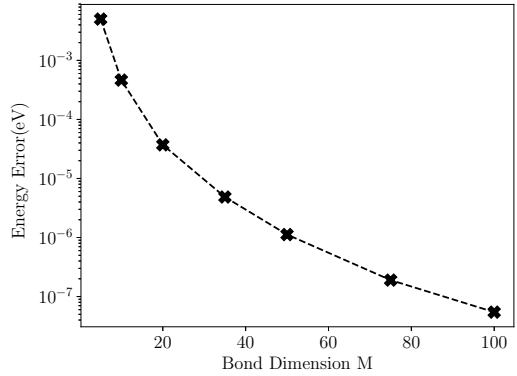}

    \caption{\label{fig:EvsM}
      The energy error of the lowest state in the one-exciton space as a function of bond dimension $M$. (The $M=500$
      energy is taken as the reference).
        }
    \end{figure}

      With the techniques described in Section \ref{sec:techniques}, the error of TD-DMRG for the zero temperature emission spectrum only
      comes from the DMRG calculation of the lowest state in the one-exciton space. The energy error of this single state is plotted as a
      function of the bond dimension $M$ in Figure \ref{fig:EvsM} (the energy calculated with $M=500$ is here regarded as exact). The single
      state DMRG calculation in the Holstein model is very accurate, even though the interaction topology in our problem is not one
      dimensional.  Unlike in purely electronic systems, where $M \sim$ several thousands is often necessary to obtain a converged ground
      state, in this electron-phonon system, $M=20$ is already enough to obtain a very accurate result with an error of less than $10^{-4}$
      eV, even though the number of sites is very large (270).

    We show the calculated zero temperature fluorescence spectrum in Figure 
    \ref{fig:DSB}(c), and the experimental spectrum at 1.4K is also plotted for
    comparison.\cite{wu2003optical} \
    The spectra calculated with $M=100$ and $M=20$ are indistinguishable. Thus, not only a single point state but also the
    spectrum can be accurately obtained with a small $M$.
    The TD-DMRG spectrum has three dominant fluorescence bands, 0-1, 0-2, 0-3;
    the 0-0 band is largely suppressed due to the H-type excitonic coupling.  To
    clearly show the fine structure, the strengths of the 0-2, 0-3 band peaks
    are multiplied by 2. Compared to the experimental spectrum, most of the fine
    structure is well reproduced. Nevertheless, the frequency of the vibration
    which couples most strongly to the exciton in the quantum chemistry derivation
    of the model
    is slightly overestimated (1658.65 $\textrm{cm}^{-1}$ from the calculation,
    $\sim$1435 $\textrm{cm}^{-1}$ from experiment) and the Huang-Rhys factor is
    underestimated, probably due to the density functional approximation. We
    also see this discrepancy in the former work, where the spectrum of the DSB dimer is
    directly treated within the density functional approximation.
    \cite{wykes2015vibronic}. In the 1-particle approximation calculation, the
    positions of the main peaks are correct. However, due to the deficiencies we
    identified in the preceding section, the 0-1 band intensity is severely overestimated,
    so that the ratio of the 0-2/0-3 band intensities to the 0-1 band intensity
    is underestimated. Though the 2-particle approximation corrects this
    problem, the number of vibrational modes which can be included in such
    calculations is too small to capture the fine structure seen in the TD-DMRG
    spectrum and in experiment.  This demonstrates the superiority of the
    TD-DMRG methods in systems with large Hilbert spaces which arise when many
    vibrations need to be considered.

\section{Conclusions}
\label{sec:conclusions}
    In this work, we implemented the zero-temperature and finite-temperature
    time-dependent DMRG algorithms (TD-DMRG) for the Holstein Hamiltonian, to
    calculate the linear absorption and fluorescence spectra of molecular
    aggregates at both zero and finite temperature.  Our calculations on PBI
    molecular chains showed that the practical accuracy of TD-DMRG reaches that
    of ML-MCTDH at zero temperature, and further allows us to extract accurate
    finite temperature dynamic properties.  The comparison with n-particle
    approximation methods on both models and the DSB crystal further shows that
    TD-DMRG is not only much more accurate than these approximations, but can
    also practically handle the larger Hilbert spaces arising from increasing
    the number of vibrational modes to model detailed spectral features.  In
    summary, our results support the use of TD-DMRG algorithms as  accurate,
    efficient and robust methods for dynamical problems including both
    electrons(excitons) and nuclei(phonons). In future work, we will carry out
    further studies using these TD-DMRG algorithms, including studies of true
    non-equilibrium phenomena, including charge and energy transport.
    
\appendix
\section{Appendix: MPO for the Holstein Hamiltonian} \label{appendix1}
    Multiplying out the matrices in the MPO representation of the Hamiltonian
    can be thought of as defining a recurrence relation to construct the
    Hamiltonian with all terms up to site $k$, from the Hamiltonian and
    operators up to site $k-1$, and the operators acting on site $k$.
    \begin{gather}
        \mathbf{\Omega}_k =
        \mathbf{\Omega}_{k-1} \cdot \textrm{MPO}_k  \\
        \mathbf{\Omega}_k =
        \begin{bmatrix}
            H_{k} & \mathbf{O}_{k} & 1 
        \end{bmatrix}
    \end{gather}
    Here, $\mathbf{\Omega}_{k}$ is the MPO obtained by
    multiplying the local MPO matrices from site 1 to $k$. $H_{k}$
    represents the part of the Hamiltonian that acts on sites 1 to $k$, while
    $\mathbf{O}_{k}$ represents the row of operators that define an
    interaction between sites 1 to $k$ and the remaining sites.  
    The general structure of the local MPO matrix of site $k$ is
    \begin{gather}
        \textrm{MPO}_k = 
    \begin{bmatrix}
        1 & \mathbf{0} & 0  \\
        \mathbf{A}_k & \mathbf{C}_k & \mathbf{0}  \\
        B_k & \mathbf{D}_k & 1  \\
    \end{bmatrix} 
    \end{gather}
    $\mathbf{A}_k$ is a column of operators, $B_k$ is a single operator,
    $\mathbf{C}_k$ is a matrix of operators, and $\mathbf{D}_k$ is a row of
    operators. The recurrence relation for each component of
    $\mathbf{\Omega}_{k}$ is
    \begin{gather}
        H_{k} = H_{k-1} \cdot 1 + \mathbf{O}_{k-1} \cdot \mathbf{A}_k + 1 \cdot  B_k \nonumber \\
        \mathbf{O}_{k} = \mathbf{O}_{k-1} \cdot \mathbf{C}_k +  1 \cdot
        \mathbf{D}_k 
    \end{gather}

    Given the order of sites in the MPS is $e_1$, $\nu_{11}$, ..., $\nu_{1n}$
    ,..., $e_i$, $\nu_{i1}$, ..., $\nu_{in}$,... , one of the optimal MPO
    representation of the Hamiltonian in eq. \eqref{eq:discreteH} is
    \begin{enumerate}[label=(\roman*)]
    
    \item For an electronic site before the middle site. \label{itm:item1}
    
    \begin{gather}
        \mathbf{A}_k = 
    \begin{bmatrix}
        J_{0k} a_k  \\
        J_{0k} a^\dagger_k \\
        ...  \\
        J_{k-1,k} a_k  \\
        J_{k-1,k} a^\dagger_k \\
    \end{bmatrix}
    \end{gather}
        
    \begin{gather}
    B_k = 
    \begin{bmatrix}
        \varepsilon_k a^\dagger_k a_k \\
    \end{bmatrix}
    \end{gather}
    
    \begin{gather}
        \mathbf{C}_k = 
    \begin{bmatrix}
        \mathbf{0} & \mathbf{I} & \mathbf{0} & \mathbf{0} \\ 
    \end{bmatrix}
    \end{gather}
    
    \begin{gather}
        \mathbf{D}_k = 
    \begin{bmatrix}
        a^\dagger_k a_k &  \mathbf{0} & a^\dagger_k & a_k  \\ 
    \end{bmatrix}
    \end{gather}
    
\item An electronic site in the middle of the MPS. $\mathbf{A}_k$ and
    $\mathbf{B}_k$ are the same as for case
      \ref{itm:item1} \label{itm:item2}
    \begin{gather}
        \mathbf{C}_k = 
    \begin{bmatrix}
        0 & J_{0,n}   &  0       & ...  \\
        0 & 0         &  J_{0,n} & ...  \\
        0 & J_{1,n}   &  0       & ...  \\
        0 & 0         &  J_{1,n} & ...  \\
       ... & ...       &  ...     & ...  \\
        0 & J_{k-1,n} &  0       & ...  \\
        0 & 0         &  J_{k-1,n} & ... \\
    \end{bmatrix}
    \end{gather}
    
    \begin{gather}
        \mathbf{D}_k = 
    \begin{bmatrix}
        a_k^\dagger a_k & J_{k,n}a^\dagger_k  & J_{k,n}a_k & ... \\ 
    \end{bmatrix}
    \end{gather}

    Multiplying out the matrices up to the middle yields a row of operators
            including complementary operators. 
    \begin{gather}
        \mathbf{\Omega}_k = 
    \begin{bmatrix}
        H_k & a_k^\dagger a_k & \textrm{Com}_n^\dagger & \textrm{Com}_n & ... &
        \textrm{Com}_{k+1}^\dagger & \textrm{Com}_{k+1} & 1\\ 
    \end{bmatrix}
    \end{gather}
    The corresponding complementary operators constructed are
    \begin{gather}
        \textrm{Com}_i^\dagger = \sum_{j=1}^{k} J_{i,j} a_j^\dagger \\
        \textrm{Com}_i = \sum_{j=1}^{k} J_{i,j} a_j
    \end{gather}
    
\item Electronic site after the middle. $B_k$, $\mathbf{D}_k$ are the same as for
      case \ref{itm:item2}
    
    \begin{gather}
        \mathbf{A}_k = 
    \begin{bmatrix}
        \mathbf{0} \\
        a_k  \\
        a^\dagger_k \\
    \end{bmatrix}
    \end{gather}
    
    \begin{gather}
        \mathbf{C}_k = 
    \begin{bmatrix}
        \mathbf{0} & \mathbf{I}  \\
    \end{bmatrix}
    \end{gather}
    
\item Vibrational site.
    \begin{gather}
        \mathbf{A}_k = 
    \begin{bmatrix}
        g_{in} \omega_{in}(b^\dagger_{in}+b_{in})  \\
        \mathbf{0}
    \end{bmatrix}
    \end{gather}
    
    \begin{gather}
     B_k = 
    \begin{bmatrix}
        \omega_{in} b^\dagger_{in}b_{in}   \\
    \end{bmatrix}
    \end{gather}
    
    \begin{gather}
        \mathbf{C}_k = 
    \begin{bmatrix}
        \mathbf{I} \\ 
    \end{bmatrix}
    \end{gather}
    
    \begin{gather}
        \mathbf{D}_k = 
    \begin{bmatrix}
        \mathbf{0} \\ 
    \end{bmatrix}
\end{gather}
    \end{enumerate}
    When excitonic coupling terms are 1-dimensional nearest-neighbor, the
    maximal bond dimension $D$ of MPO constructed above is $\sim \textrm{const}$,
    otherwise $D \sim k_e$. 

\begin{acknowledgement}
    Support for JR and ZS is from the National Natural Science Foundation of
    China via Science Center for Luminescence from Molecular Aggregates (CELMA)
    (Grant no.21788102) and Ministry of Science and Technology of the People's
    Republic of China (Grant no.2017YFA0204501). Support for GKC is from the US
    National Science Foundation via grant no. CHE-1665333. JR is grateful to Dr.
    Zhendong Li, Dr. Enrico Ronca and Chong Sun for stimulating discussions, to
    Dr. Wenqiang Li for providing the quantum chemistry parameters of the DSB
    crystal.
\end{acknowledgement}

\begin{suppinfo}
    \begin{itemize}
        \item Filename: SI.pdf \\
           Stability analysis of the RK4 method, the comparison of the star/chain
            representations in TD-DMRG, and the quantum chemistry parameters in
            the DSB crystal.
    \end{itemize}
    This information is available free of charge via the Internet at http://pubs.acs.org
\end{suppinfo}

\bibliography{ephMPS}

\end{document}